\documentclass[aps,prl,floatfix,amsmath,amssymb,showpacs,unsortedaddress,superscriptaddress,twocolumn]{revtex4-1}\usepackage{amsmath, amssymb, graphics, mathrsfs, bm, stackrel,bbold}
\usepackage{graphicx}

\usepackage[usenames,dvipsnames]{xcolor}%http://en.wikibooks.org/wiki/LaTeX/Colors
\definecolor{Highlight}{rgb}{1,1,0.75}

\allowdisplaybreaks[1]
\usepackage{graphicx}
\usepackage[pdftex, bookmarks={false}, pdfauthor={Srividya Iyer-Biswas}, pdftitle={Title}]{hyperref}
\hypersetup{colorlinks=true, linkcolor=BrickRed, citecolor=Violet, filecolor=OliveGreen, urlcolor=RoyalBlue, filebordercolor={.8 .8 1}, urlbordercolor={.8 .8 0}}%http://en.wikibooks.org/wiki/LaTeX/Hyperlinks
\usepackage[normalem]{ulem}					% add \sout strickout

\newcommand\ba{\begin{array}}
\newcommand\ea{\end{array}}
\newcommand\nn{\nonumber}
\newcommand\ri{\right}
\renewcommand\le{\left}

\newcommand{\feyn}[1]{#1\kern-0.45em/}

\newcommand{\tto}{\rightarrow}

\renewcommand\a{\alpha}%metric symbol

\renewcommand\b{\beta}%metric symbol

\renewcommand\d{\delta}%accent

\newcommand\G{\Gamma}

\renewcommand\k{\kappa}%accent

\renewcommand\l{\lambda}%non-ascii letter
\newcommand\m{\mu}

\newcommand\s{\sigma}

\renewcommand\t{\tau}%accent char

\renewcommand\th{\theta}%latin char

\newcommand\x{\xi}

\newcommand\X{\Xi}

\newcommand\la{\langle}
\newcommand\ra{\rangle}
\newcommand\pd{\partial}

\newcommand\mb{\mathbb}
\newcommand\mbs{\boldsymbol}

\newcommand\ms{\mathscr}

\newcommand{\eqwithrate}[1]{\mathrel{\mathop{\longrightarrow}\limits^{#1}}}

\begin{document}
\title{Universality in stochastic exponential growth}
\author{Srividya Iyer-Biswas}
\affiliation{James Franck Institute and Institute for Biophysical Dynamics,  University of Chicago, Chicago, IL 60637}
%\email{iyerbiswas@uchicago..edu}
\author{Gavin E. Crooks}
\affiliation{Physical Biosciences Division, Lawrence Berkeley National Laboratory, Berkeley, CA 94720}
\author{Norbert F. Scherer}
\email{nfschere@uchicago.edu}
\affiliation{James Franck Institute and Institute for Biophysical Dynamics,  University of Chicago, Chicago, IL 60637}
\author{Aaron R. Dinner}
\email{dinner@uchicago.edu}
\affiliation{James Franck Institute and Institute for Biophysical Dynamics,  University of Chicago, Chicago, IL 60637}

%\date{\today}
\begin{abstract}
Recent imaging data for single bacterial cells reveal that their mean sizes grow exponentially in time and that their size distributions collapse to a single curve when rescaled by their means. An analogous result holds for the division-time distributions. A model is needed to delineate the minimal requirements for these scaling behaviors. We formulate a microscopic theory of stochastic exponential growth as a Master Equation that accounts for these observations, in contrast to existing quantitative models of stochastic exponential growth (e.g., the Black-Scholes equation or geometric Brownian motion). Our model, the stochastic Hinshelwood cycle (SHC), is an autocatalytic reaction cycle in which each molecular species catalyzes the production of the next. By finding exact analytical solutions to the SHC and the corresponding first passage time problem, we uncover universal signatures of fluctuations in exponential growth and division. The model makes minimal assumptions, and we describe how more complex reaction networks can reduce to such a cycle. We thus expect similar scalings to be discovered in stochastic processes resulting in exponential growth that appear in diverse contexts such as cosmology, finance, technology, and population growth.
\end{abstract}
\maketitle

Discovering unifying physical principles that transcend the complexity of 
specific biological systems is a fundamental goal of the field of biological physics~\cite{2012-bialek-fk, 1970-delbruck-hl}. Quantitative analyses of gene regulatory networks have revealed general connections between network motifs, fluctuations in the dynamics of participating molecules, and biological functions {at the molecular scale}~\cite{Munsky2012, 2006-friedman-qd, 2012-salman-cr, Maienschein2010}. Analogous quantitative relationships governing behaviors at the organismal scale are just beginning to emerge \cite{2011-scott-kx, 2012-velenich-vn, 2012-bialek-fk}.  In particular, in a recent experiment, we found that scaling laws governed the stochastic growth of individual {\em Caulobacter crescentus} cells~\cite{-iyer-biswas-kv}. In the same study, the sizes of the cells were shown to increase {\em exponentially} between divisions, consistent with observations for other microorganisms \cite{2012-chien-mz, 2010-hagen-fk,  2011-mir-qf, 2007-di-talia-fv}.

Exponential growth is ubiquitous and has been studied in diverse contexts~\cite{2010-hagen-fk, 1949-monod-uq}.  It describes inflation of the universe, geometric multiplication of an entity of interest (e.g., nuclear or cellular fission),  and phenomenological dynamics (e.g., the Black-Scholes equation for options prices; Moore's Law for computer processor power). Many such processes are inherently stochastic, with the times between contributing events drawn from waiting-time distributions~\cite{2008-feller-fk}.  Surprisingly given its prevalence, there is no microscopic model for stochastic exponential growth. While various other physical aspects of cell growth have been examined previously~\cite{2011-deng-rz, 2007-goyal-kq, 2010-jiang-ly, 2012-furusawa-eu, 2011-goldstein-pd}, a  theory relating the statistics of the stochastic exponential growth to essential features of the biochemical networks underlying growth is needed.

{A phenomenological model of stochastic exponential growth, a Langevin equation with linear drift and linear multiplicative noise,  was famously applied by Black-Scholes to explain financial data on stock options prices; it forms the basis of modern quantitative derivative trading~\cite{1975-black-qf}. This model, also known as Geometric Brownian Motion (GBM), has since been used extensively in various cellular contexts, and when applied to cell growth, it predicts a lognormal cell size distribution~\cite{2005-furusawa-rr}. However, in this model, the standard deviation grows faster than the exponentially growing mean such that the ratio, i.e., the coefficient of variation (COV), increases as the square-root of time. This prediction is in disagreement with observations in \cite{-iyer-biswas-kv}, wherein the COV of cell sizes was found to be time invariant. }

Here, we provide a microscopic theory of stochastic exponential growth that yields the universality of fluctuations during the growth of single bacterial cells, observed in \cite{-iyer-biswas-kv}; it also agrees with the aforementioned observed constancy of the COV with time.  This microscopic theory is built on the assumption that {growth} is governed by an autocatalytic cycle of reactions. We argue {\em a posteriori} that this is the minimal model consistent with observations. Furthermore, we provide a theoretical framework for examining stochastic cell division and show how scale invariance of division time distributions arises. We also discuss why the essential features of the model are retained even when more complex network topologies govern cell growth.

%%%%%%%%%%%%%%%%%%%%%%begin%%%%%%%%%%%%%%%%%%%%%%
\begin{figure}[bt]
\begin{center}
\resizebox{3.25in}{!}{\includegraphics{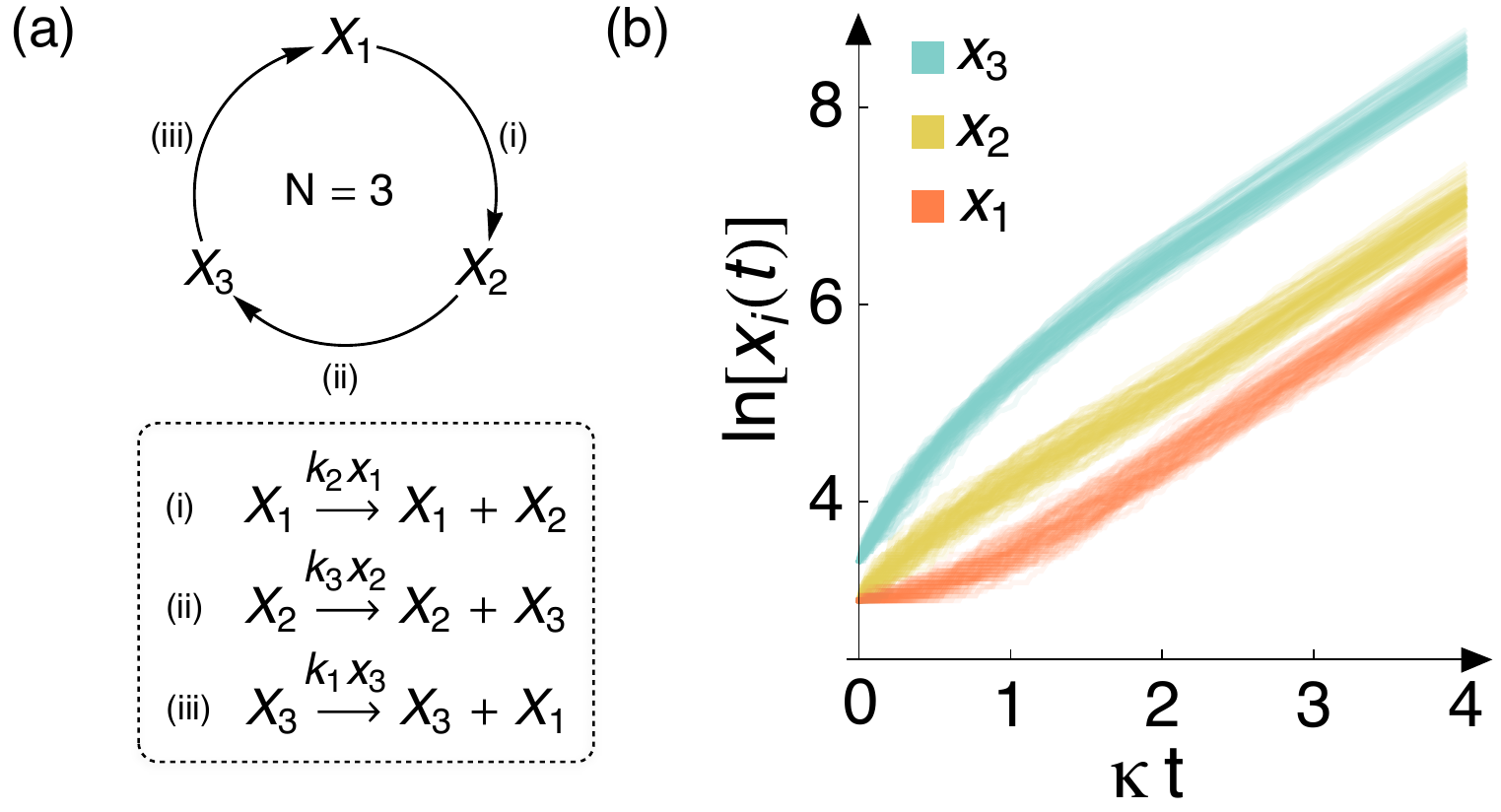}}%inclgphcs[trim=lcm bcm rcm tcm, clip=true, angle=-90]
\caption{{\bf Stochastic Hinshelwood cycle (SHC).  (a)} Schematic of the cycle.  An example with three chemical species ($N=3$) is shown.  The curved arrows indicate that the production of each species, $X_i$, is catalyzed by the previous one, $X_{i-1}$,  with rate  $k_ix_{i-1}$, where $x_{i-1}$ is the copy number of $X_{i-1}$. The box shows the corresponding reactions explicitly (see \eqref{eq-model} for the general specification with $N$ species).  {\bf (b)} Stochastic exponential growth trajectories for the model shown in (a).  A single composite timescale emerges asymptotically:  $x_{1}$, $x_{2}$,  and $x_{3}$ all grow with the same exponential growth rate, $\k = (k_{1}k_{2}k_{3})^{1/3}$, which is the mean slope for each curve in the log-linear plot, for $t \gg 1/\k$. We show the evolution of the three species for 100 stochastic trajectories.  They are obtained from Gillespie simulations \cite{Gillespie1977} of \eqref{eq-cme} for rate constants $\mbs{k} = (0.1, 1.5, 3.2)$ and initial copy numbers $\mbs{x}(0)= (20, 20, 30)$. %$\k$ is found to be $0.78$.
}
\label{fig-SHC}
\end{center}
\end{figure}
%%%%%%%%%%%%%%%%%%%%%%%end%%%%%%%%%%%%%%%%%%%%%%

{\em  Stochastic Hinshelwood  cycle:}  
Our stochastic theory builds on a simple deterministic (kinetic) model introduced in 1952 by Hinshelwood~\cite{1952-hinshelwood-fk}. In this model  components of the cell that govern cell growth are  connected through an autocatalytic cycle of reactions in which each species catalyzes the production of the next  (Fig.\ \ref{fig-SHC}a). The mass (or equivalently, the size) of a cell is assumed to be proportional to a linear combination of the copy numbers of the species in the cycle. We construct a stochastic generalization of this cycle, by assuming that the waiting times for the individual reactions are exponentially distributed, i.e., that the reactions are elementary. We refer to this model as the {\em stochastic} Hinshelwood cycle (SHC).  In general, the SHC contains $N$  species, $\{X_{1}, X_{2}, \ldots, X_{N}\}$.  The scaling laws that we  derive do not depend on their identities or $N$.   The mean rate of production of $X_i$  is  $k_ix_{i-1}$, where $x_{i-1}$ is the copy number of $X_{i-1}$ (Fig.\ \ref{fig-SHC}).  For use below, we write this rate as a matrix multiplication:  $k_ix_{i-1}=\sum_{j=1}^N\mb{K}_{ij}\, x_{j}$, where $\mb{K}$ is the rate constant matrix with elements $\mb{K}_{ij} = k_{i}\,\d_{i-1, j}$ and $\d$ is the Kronecker delta. In this notation, the reaction scheme is 
%%%%%%%%%%%%%%%%%%%%%%begin%%%%%%%%%%%%%%%%%%%%%%
\begin{align}
\label{eq-model}
X_{i-1} \eqwithrate{\sum_{j=1}^N\mb{K}_{ij}\, x_{j}} X_{i-1} + X_{i},
\end{align}
%
%%%%%%%%%%%%%%%%%%%%%%%end%%%%%%%%%%%%%%%%%%%%%%
for $1<i,j<N$; the index 0 is equivalent to $N$, closing the cycle.

We denote the state of the general $N$-step SHC model by the vector $ \mbs{x}  \equiv (x_{1}, x_{2}, \ldots, x_{N})$, where $x_i$ is the copy number of $X_i$ present at a given time. The corresponding Chemical Master Equation (CME) \cite{2008-feller-fk,Gillespie1977} for the time evolution of the probability distribution, $P(\mbs{x}, t)$, is
%%%%%%%%%%%%%%%%%%%%%%begin%%%%%%%%%%%%%%%%%%%%%%
\begin{align}
\label{eq-cme}
\frac{\pd P}{\pd t} =\sum_{i,j=1}^N \mb{K}_{ij}\, x_{j}\bigl[P(.., x_{i}-1, ...) - P(.., x_{i}, ...) \bigr]. 
\end{align}
%%%%%%%%%%%%%%%%%%%%%%%end%%%%%%%%%%%%%%%%%%%%%%
From \eqref{eq-cme}, we  derive the time evolution equations for the moments of $\mbs{x}$ from the eigenvalues and eigenvectors for $\mb{K}$. Since $\mb{K}$ is a cyclic matrix of period $N$,  $\mb{K}^{N} =  k_{1}\,k_{2}\,\ldots k_{N}\,\mb{1} \equiv \k^{N}\, \mb{1}$ \cite{2008-feller-fk}, and the eigenvalues of the rate constant matrix are the $N$ complex roots of unity times $\k$, the geometric mean of all the rates.  The $m^{{th}}$ eigenvalue is $\l_{m} = \k\,\exp \le({i\,  2\pi m/N}\ri)$, and the  $q^{{th}}$ component of the corresponding eigenvector is ${\x}_m^{(q)}={\le( \prod_{p=1}^{q}k_{p}\ri)} /{\l_{m}^{q}}$. $\l_{N}=\k$ is the eigenvalue with the largest positive real part; thus the time scale that dominates the asymptotic dynamics is $\k^{-1}$.

{\em Time evolution of the mean copy numbers:}  The CME  dictates that the (ensemble averaged) mean copy numbers of the reactants, $\mbs{\m}(t)$, evolve with time according to $d\mbs{{\m}}/dt = \mb{K}\mbs{\m}(t)$ \cite{2008-feller-fk}. The formal solution to this equation is $\mbs{\m}(t) = \exp({\mb{K\,}t})\mbs{\m}(0)$, or equivalently $\m_{m} (t) = \sum_{i,j=1}^N\mb{U}_{mi} \,e^{\l_{i}t} \,\mb{U}^{-1}_{ij}\,\, \m_{j}(0)$, where $\mb{U}$ is the matrix of eigenvectors $\mb{U} = \le[ \mbs{\x}_{1}\,\,\, \mbs{\x}_{2} \,\,\, \ldots \,\,\mbs{\x}_{N}\ri]$ \cite{2008-feller-fk}.  In the asymptotic time limit (i.e., when $t \gg 1/\k$), 
%%%%%%%%%%%%%%%%%%%%%%begin%%%%%%%%%%%%%%%%%%%%%%
\begin{align}\label{eq-means}
\m_{q} (t) \sim \sum_{i=1}^N \mb{U}_{qN}  \,\mb{U}_{Ni}^{-1}\,{\m_{i}(0)}\, e^{\k t}.
\end{align}
%%%%%%%%%%%%%%%%%%%%%%%end%%%%%%%%%%%%%%%%%%%%%%
Thus the mean copy numbers of all reactants evolve asymptotically as $e^{\k t}$.  Moreover,  the dependence on initial conditions for $\m_{q} (t)$ is independent of $q$.  It follows that the ratio of any two mean copy numbers, $\m_{q} (t)/ \m_{r} (t)$, is equal to $\mb{U}_{q\, N}/ \mb{U}_{r\, N}$, which is {\em independent of initial conditions} and depends only on the $q^{{th}}$ and $r^{{th}}$ components of the $N^{{th}}$ eigenvector, $\mbs{\x}_{N}$.

{\em Time evolution of growth fluctuations:} To examine the time evolution of growth fluctuations, we determine the equation of motion of the  covariance matrix, $\mb{{
C}}_{ij} \equiv \le[ \la x_{i} \, x_{j}\ra- \la x_{i} \ra\, \la x_{j}\ra \ri] $ \cite{2008-feller-fk}.  
In matrix form,
%%%%%%%%%%%%%%%%%%%%%%begin%%%%%%%%%%%%%%%%%%%%%%
\begin{align}%\label{eq-}
\frac{d}{dt}\mb{{C}}(t) = \mb{K}\, \mb{C}(t) + \mb{C}(t)\, \mb{K}^{\intercal} + \frac{d}{dt}{\mb{\X}}(t),
\end{align}
%%%%%%%%%%%%%%%%%%%%%%%end%%%%%%%%%%%%%%%%%%%%%%
where $\intercal$ denotes the transpose and $\mb{\X}(t)$ is an $N \times N$ diagonal matrix with  entries $\mb{\X}_{ij}(t) = \d_{ij}\, \m_{j}(t)$. We have computed the exact analytical solution for the time evolution of the covariance matrix \cite{bnote}.
In the asymptotic limit, 
%%%%%%%%%%%%%%%%%%%%%%begin%%%%%%%%%%%%%%%%%%%%%%
\begin{align}\label{eq-cov1}
\mb{C}_{ij}\,(t) \sim\mb{U}_{iN} \,\mb{U}_{jN} \, e^{2 \k t} \, \sum_{p=1}^N{b}_{p} \,\ms{\m}_{p}\,(0), 
\end{align}
%%%%%%%%%%%%%%%%%%%%%%%end%%%%%%%%%%%%%%%%%%%%%%
where $b_{p}$ is a coefficient that  depends only on the rates and not the initial conditions \cite{bnote}.  
Thus, $\mb{C}_{ij}$ scales as $e^{2 \k t}$ for all $i$ and $j$. Moreover, the time-independent pre-factor of  element  $\mb{C}_{ij}$ of the covariance matrix is proportional to $\mb{U}_{i N} \,\mb{U}_{j N}$.  
Combining \eqref{eq-cov1} with \eqref{eq-means} gives 
%%%%%%%%%%%%%%%%%%%%%%begin%%%%%%%%%%%%%%%%%%%%%%
\begin{align}
\label{eq-cov}
\mbox{Cov}\le[{x}_{i}(t)/\m_{i}(t), {x}_{j}(t)/\m_{j}(t)\ri]/\s_{i}\,\s_{j} \sim 1,
\end{align}
%%%%%%%%%%%%%%%%%%%%%%%end%%%%%%%%%%%%%%%%%%%%%%
where $\sigma_{i}$ is the standard deviation of the rescaled variable $x_{i}/\m_{i}(t) $. 
%%%%%%%%%%%%%%%%%%%%%%begin%%%%%%%%%%%%%%%%%%%%%%
\begin{figure}[bt]
\begin{center}
\resizebox{3.25in}{!}{\includegraphics{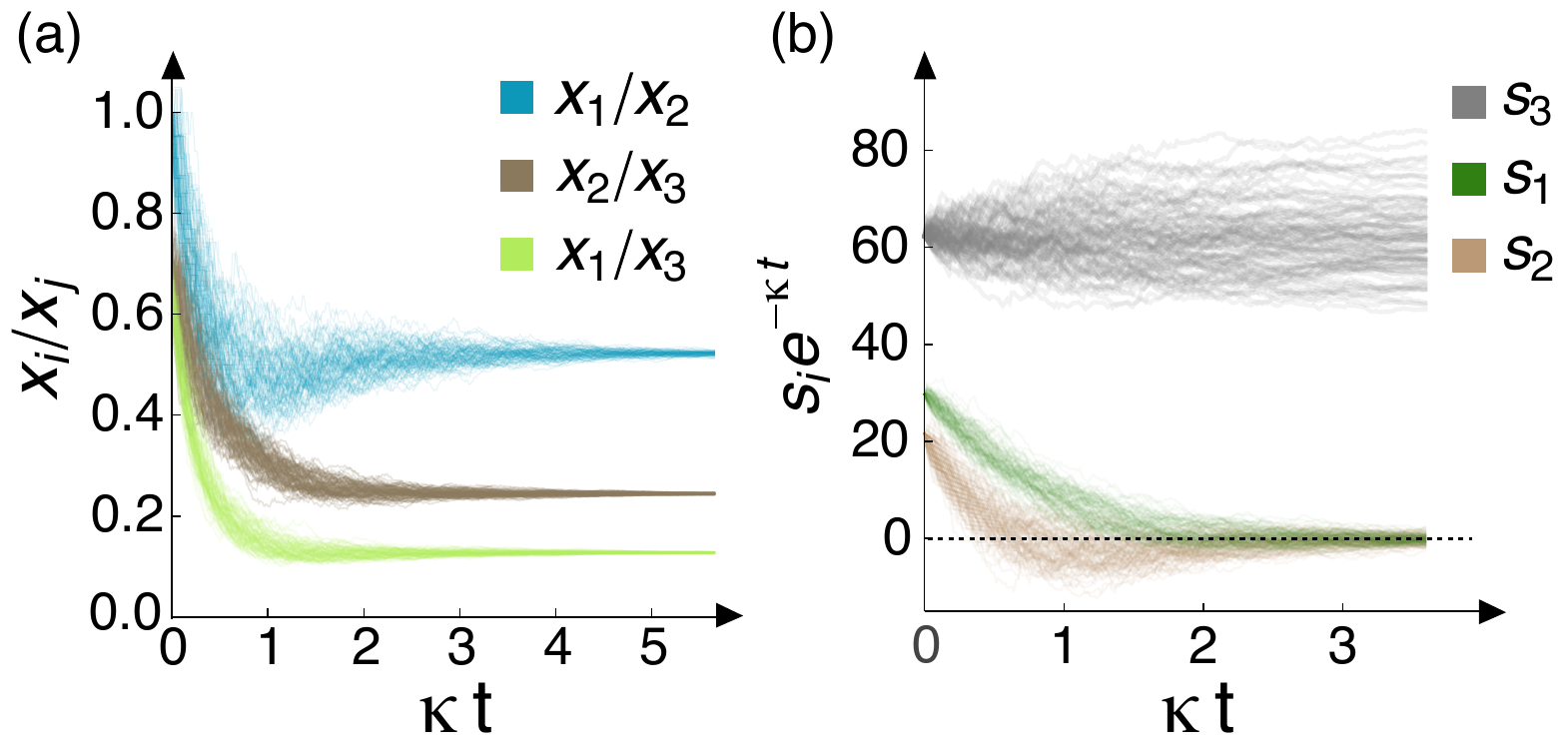}}%inclgphcs[trim=lcm bcm rcm tcm, clip=true, angle=-90]
\caption{{\bf Copy number fluctuations are perfectly correlated in the asymptotic limit. (a)} Ratios of the copy numbers of the components of the $N=3$ SHC from the  trajectories  shown in Fig.~\ref{fig-SHC}b.  As predicted by \eqref{eq-cov}, for $t \gg 1/\k$, the ratios of the different $x_{i}$ tend to constant values in each ensemble member, a signature of the perfect correlations between component copy numbers in the asymptotic state. {\bf (b)} Emergence of a single composite variable.  The variable $s_i$ is the projection of the state vector $\mbs{x}$ onto the $i^{th}$ eigenvector ($\mbs{\x}_i$) of the rate constant matrix, $\mb{K}$.  We see that $s_3$ here (or more generally, $s_N$) tends to a constant non-zero level, while the remainder of the projections vanish.  %Thus, all $x_{i}(t)$ are proportional to $s_{3}(t)$ and become perfectly correlated to each other. 
}
\label{fig-pc}
\end{center}
\end{figure}
%%%%%%%%%%%%%%%%%%%%%%%end%%%%%%%%%%%%%%%%%%%%%%

An important consequence of \eqref{eq-cov} is that asymptotically all $x_{i}$ are proportional to each other, since two stochastic variables can be perfectly correlated only when they are linearly related  \cite{2008-feller-fk}. Thus, the ratio of any two of them must asymptote to a time-independent constant value in each ensemble member (i.e., each cell; see Fig.~\ref{fig-pc}a), but this value itself has a distribution across different  members. We note that $x_{i}(t)$ and $x_{j}(t)$ themselves continue to fluctuate in each stochastic realization even as their ratio tends to a constant value.

{\em Scalings of the size distribution:}   
Two different scaling laws are encapsulated in  \eqref{eq-cov}. First, every rescaled variable  $x_{i}/\m_{i}(t) $ has the same distribution in the asymptotic limit. Second, since $e^{\k t}$ is a scaling variable, the distribution shape for each $x_{i}$ is invariant with time, even as its mean increases exponentially. Therefore, the $n^{{th}}$ moment of $x_{i}$ goes as $e^{n\k t}$. 

For clarity, we explicitly compute the size distribution for the case when all rate constants in the model are equal, with value $\k$.  In this case, $\mb{K}$ becomes a circulant matrix, and the projection of the state vector $\mbs{x}$ onto the asymptotically dominant eigenvector ${\mbs\x}_N$ reduces to a simple sum of the constituent copy numbers, $s \equiv  \sum_{i=1}^{N} x_{i}$. This variable, $s$, itself undergoes dynamics governed by a $N=1$ SHC.  Then, $P(s,t)$, for the initial condition $P(s,t=0) = \d_{s,s_{0}}$, is the negative binomial distribution,
%%%%%%%%%%%%%%%%%%%%%%begin%%%%%%%%%%%%%%%%%%%%%%
\begin{align}\label{eq-nbin}
P(s,t| s_{0}, 0) =  \binom{s-1}{s_{0}-1}  \left(e^{-\k t}\right)^{s_0} \left(1-e^{-\k t}\right)^{s-s_0}.
\end{align}
%%%%%%%%%%%%%%%%%%%%%%%end%%%%%%%%%%%%%%%%%%%%%%
This result can be verified by direct substitution into  \eqref{eq-cme}. In the continuum limit for $s$, \eqref{eq-nbin} tends to a gamma distribution, since the negative binomial distribution can be written as a Poisson mixture of gamma distributions \cite{Karlis2005}. Asymptotically,
%%%%%%%%%%%%%%%%%%%%%%begin%%%%%%%%%%%%%%%%%%%%%%
\begin{align}
\label{eq-gamma}
P(s, t \tto \infty|s_{0},0) = \frac{ s^{s_{0}-1}e^{-\le(s_{0} \, s\ri)}}{s_{0}^{-s_{0}}\, {\G}(s_{0})}.
\end{align}
For the general case with unequal rates, the analog of $s$ is the linear combination of $\{x_i\}$ that is defined by the projection of the state vector along the eigenvector corresponding to the largest eigenvalue, $\k$:  $s_N\equiv \sum_{i=1}^N\mb{U}^{-1}_{Ni}x_i$.  As shown in Fig.~\ref{fig-pc}b, all $s_{i} \equiv \sum_{j=1}^N\mb{U}^{-1}_{ij} x_{j}$ for $i \neq N$ vanish in the long-time limit, and the only contributions to fluctuations in each $x_i$ come from $s_N$.  As a result, all ${x}_i/\m_{i}$ are distributed in the same fashion as $s$ in \eqref{eq-gamma} (Fig.~\ref{fig-growth-sc}), with $s_{0} = s_{N}(0)$.   In other words, the mean-rescaled distribution of cell sizes must fit the same gamma distribution at all times.

{\em Division as the first passage time (FPT) to a size threshold:} 
We assume that cell division occurs when the cell size, $s$, reaches a threshold \cite{2007-di-talia-fv,-iyer-biswas-kv}. In general, this threshold can be absolute ($s$ itself attains a critical value), relative ($s$ increases by a critical multiple), or differential ($s$ increases by a critical amount). In the absence of additional feedback mechanisms, the scaling derived above implies that the different components of the SHC maintain their pre-division ratios, not just in the mean, but also in their fluctuations. Moreover, the thresholding prescription (absolute, relative, or differential) can be applied to any one component of the SHC. All remaining components of the SHC, as well as the total size, will simply follow, because they are perfectly correlated.

For an absolute size threshold,  the first passage time distribution, $\mathcal{P}(\t)$, can be determined from the fraction of trajectories crossing the threshold ($\theta$) in a given time interval ($\tau$ to $\tau+\Delta\tau$), by utilizing the fact that the stochastic size variable is monotonically increasing:
%%%%%%%%%%%%%%%%%%%%%%begin%%%%%%%%%%%%%%%%%%%%%%
\begin{align}
\label{eq-fpt-gen}
\mathcal{P}(\t) = \frac{\pd}{\pd \t} \le[ \int_{\th}^{{\infty}} ds \,P(s, \t) \ri].
\end{align}
%%%%%%%%%%%%%%%%%%%%%%%end%%%%%%%%%%%%%%%%%%%%%%
Substituting \eqref{eq-gamma} into \eqref{eq-fpt-gen}, we find that the first passage time distribution from a given initial size $s_0$ to an absolute threshold $\theta$ is a beta-exponential distribution \cite{2006-nadarajah-ly},
%%%%%%%%%%%%%%%%%%%%%%begin%%%%%%%%%%%%%%%%%%%%%%
\begin{align}
\label{eq-fpt}
\mathcal{P}(\t) = \frac{\k\,e^{-s_0 \k \tau } \left(1-e^{-\k \tau }\right)^{\theta -s_0}}{{\rm B}(s_0,1+ \th-s_{0})},
\end{align}
%%%%%%%%%%%%%%%%%%%%%%%end%%%%%%%%%%%%%%%%%%%%%%
where ${\rm B}$ is the beta function. The FPT distributions for differential or relative size thresholds can be found using this expression.  

{\em Scalings of division times:}
Since $\t$ always occurs as  $\k \t$  in \eqref{eq-fpt}, we can rescale time by $\k^{-1}$, or, equivalently, by $\la \t \ra$,  to obtain a universal scale-invariant shape of the division time distribution. A complementary translational collapse of $\mathcal{P}(\t)$ is obtained when  $\t$ is shifted to $\t - \log(\th/\k)$, provided that $\th \gg s_{0}$. The scale-invariance of first passage time distributions is more universal than \eqref{eq-fpt}; a similar scaling collapse of the division time distribution will be obtained whenever a single timescale dominates the dynamics, regardless of the thresholding scheme (i.e., absolute, relative, or differential). Operationally, this implies that if $\k$ is  varied by changing an external parameter (e.g., nutrient quality, oxygen concentration, osmotic pressure or temperature), the mean-rescaled division time distributions for different values of $\k$ should collapse to the same curve, as observed in \cite{-iyer-biswas-kv}. 

{{\em Extensions of the SHC model:}
More complex autocatalytic network topologies can be specified by augmenting $\mb{K}$ in \eqref{eq-model} by additional non-zero entries. In this case, the characteristic polynomial that determines the (complex) eigenvalues, $\l$, of the augmented reaction matrix is~\cite{Iyer-Biswas:aa}
%%%%%%%%%%%%%%%%%%%%%%%%begin%%%%%%%%%%%%%%%%%%%%%%%%
\begin{align}%\label{eq-}
\sum_{\tiny{cycles}} \le( \frac{\k_{\tiny{cycle}}}{\l} \ri)^{N_{\tiny{cycle}}} = 1.
\end{align}
%%%%%%%%%%%%%%%%%%%%%%%%%end%%%%%%%%%%%%%%%%%%%%%%%%
{In other words, a complex autocatalytic network can be factorized  into irreducible cycles, each with $N_{\tiny{cycle}}$ members \footnote{See supplement.}. $\k_{\tiny{cycle}}$ is the geometric mean of rates of a given cycle. Since there is always one cycle with all $N$ members, the order of the polynomial is $N$. The largest $\k_{\tiny{cycle}}$ determines which cycle dominates the asymptotic dynamics; the linked members of that cycle specify an effective SHC, and remaining species  entrain to its stochastic exponential growth dynamics. 
Thus all the SHC scaling predictions continue to hold for more complex topologies~\cite{Iyer-Biswas:aa}. 

%%%%%%%%%%%%%%%%%%%end%%%%%%%%%%%%%%%%%%%%%%

%%%%%%%%%%%%%%%%%%%%%%%end%%%%%%%%%%%%%%%%%%%%%%
%%%%%%%%%%%%%%%%%%%%%%begin%%%%%%%%%%%%%%%%%%%%%%
\begin{figure}[bt]
\begin{center}
\resizebox{3.25in}{!}{\includegraphics[trim=.0cm 0cm 0cm 0cm, clip=true]{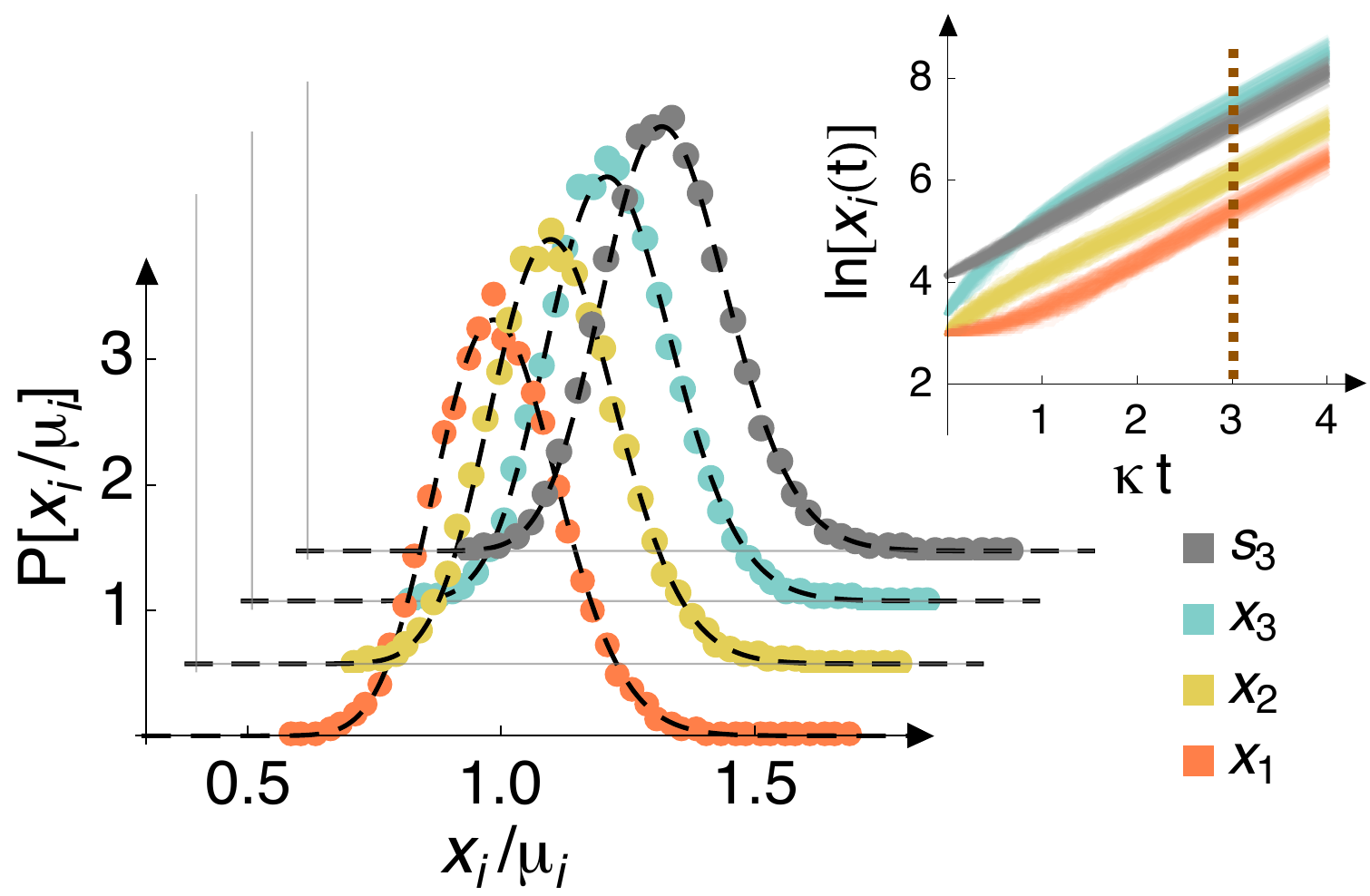}}%inclgphcs[trim=lcm bcm rcm tcm, clip=true, angle=-90]
\caption{{\bf Universality of growth fluctuations.} The symbols mark the numerically obtained distributions of copy numbers, rescaled by their means, for the trajectories shown in Figs.\ \ref{fig-SHC}b and \ref{fig-pc}, at the time indicated by the brown dotted line in the inset ($\k t = 3$).  The analogous distribution for  the composite stochastic variable,  $s_{3}$ (the projection of the state vector onto the dominant eigenvector), is also shown.  The black dashed curves are the gamma distribution in \eqref{eq-gamma}. The inset superimposes the composite stochastic variable (gray curves) on the trajectories in Fig.\ \ref{fig-SHC}b. The fact that the trajectories do not converge or diverge with time in this representation also indicates that the distributions of all the $x_{i}$ are time invariant asymptotically when these variables are rescaled by their exponentially growing means.}
\label{fig-growth-sc}
\end{center}
\end{figure}
%%%%%%%%%%%%%%%%%%%%%%%end%%%%%%%%%%%%%%%%%%%%%%

{{\it Discussion:}}
In this paper, we have introduced the stochastic Hinshelwood cycle (Fig.\ \ref{fig-SHC}), a model of stochastic exponential growth. Its dynamics naturally lead to the emergence of a single composite growth variable with a single  timescale,  thus yielding scaling collapses for size and division time distributions (Figs.~\ref{fig-pc} and \ref{fig-growth-sc}), as observed in \cite{-iyer-biswas-kv}. Moreover, this model explains the observed Arrhenius scaling of the exponential growth rate~\cite{-iyer-biswas-kv,1942-monod-kx,1958-schaechter-ly, 1979-herendeen-qf}: since the effective exponential growth rate is the {\em geometric} mean of the individual rates, the effective activation energy barrier is the {\em arithmetic} mean of the individual ones, and thus of the  order of a single enzyme reaction's, i.e., $\approx$13 kcal/mol~\cite{-iyer-biswas-kv}. 

Unlike GBM, the SHC model predicts that the ratio of the standard deviation to the mean (COV) of cell sizes is asymptotically a constant, in agreement with observations in \cite{-iyer-biswas-kv}. This can be directly seen from \eqref{eq-cov}, or by writing down the phenomenological Langevin description corresponding to the SHC: $ds/dt =  \k \,s(t) + \sqrt{s(t)}\,\eta(t)$ ($\eta$ is standard delta-correlated Gaussian white noise), whose solution is the gamma distribution in \eqref{eq-gamma}~\cite{Dornic2005,Feller1951}.  In contrast, GBM has a noise term ${s(t)}\,\eta(t)$ in the variables above and results in a lognormal size distribution \cite{2005-furusawa-rr}
\footnote{Division at a time specified up to Gaussian noise is another route to a lognormal size distribution  \cite{2013-amir-vn}.  
} with an asymptotically diverging COV ($\sim \sqrt{t}$). 

The differences in the predictions of the two models (SHC vs.\ GBM) have important implications.  In favorable chemostatic conditions, bacterial cells grow exponentially at a constant rate.  The single-cell analog of this  ``balanced growth condition'' is that the mean-rescaled cell-size distributions remain invariant, even as the cells grow and divide. This has been observed in \cite{-iyer-biswas-kv}, and is obtained from the SHC but not GBM. In the SHC the scaling collapse of cell-size distributions reflects the statistical self-similarity of the underlying stochastic process, which ensures constancy of COV.

The success of the SHC raises the question of its molecular origin. As discussed, a complex  autocatalytic network governing cell growth can be systematically reduced to an effective SHC, with an exponential growth rate determined by a subset of connections. Moreover, the mechanics of cell wall synthesis must be coupled to cell growth via the regulation of number density of active growth sites by a component of the SHC~\cite{2012-amir-xy}.  Previous studies have found indirect evidence that there are two key  steps governing bacterial growth: the global production of proteins at a rate proportional to the numbers of ribosomal RNA and vice-versa~\cite{1958-schaechter-ly, 1979-herendeen-qf}---in essence,  an $N=2$ stochastic Hinshelwood Cycle~\cite{2010-hagen-fk}. It would be interesting to test these ideas directly by designing perturbations that give rise to $N$-dependent  transients.

\begin{acknowledgments}
{\it Acknowledgments:} We thank Tom Witten and Leo Kadanoff for several insightful discussions, and Herman Gudjonson for a careful reading of the manuscript.  A.R.D., N.F.S., and S.I.B. thank the W. M. Keck Foundation and the National Science Foundation (NSF PHY-1305542) for financial support.  G.E.C. was supported by the Office of Basic Energy Sciences of the U.S. Department of Energy under Contract No. DE-AC02- 05CH11231.
\end{acknowledgments}

%\bibliography{shc}

\begin{thebibliography}{10}
\bibitem{bnote}%
  \BibitemOpen
  \bibinfo {note} {$\mb{C}=\mb{U}\mb{M}\mb{V}^{-1}$; $\mb{V} =
  (\mb{U}^\intercal)^{-1}$; $\mb{M} \equiv \mb{M}^{(1)} + \mb{M}^{(2)}$;
  $\mb{M}^{(1)}_{ij} = e^{(\l_{i} + \l_{j})t}\mb{M}^{(1)}_{ij}(0)$ and
  $\mb{M}^{(2)}_{ij}= \sum_{p,q,r=1}^N \mb{V}_{pi} \mb{V}_{pj}
  \mb{U}_{pq}\times \l_{q} \mb{V}_{rq} \mbs{\m}_{r}(0)[e^{(\l_{i} + \l_{j}) t}
  - e^{\l_{q}t}]/(\l_{i} + \l_{j} - \l_{q})$. $\mb{M}^{(2)}_{NN}(t) \sim
  \sum_{p,q,r=1}^N \mb{V}_{pN} \mb{V}_{pN} \mb{U}_{pq} \l_{q} \mb{V}_{rq}
  \mbs{\m}_{r}(0)e^{2\k t}/(2\k - \l_q)\equiv \sum_{r=1}^Ne^{2 \kappa t} b_r
  \mbs{\mu}_r(0)$, which defines $b_r$.}%
  \bibAnnoteFile{Stop}{bnote}%
\end{thebibliography}

%

%%%%%%%%%%%%%%%%%%%Supplemental Material%%%%%%%%%%%%%%%%%%%
\onecolumngrid
\pagebreak
%\vspace{1cm}
\begin{center}
{\bf\Large Supplemental Material}
\end{center}
%\vspace{0.1cm}
\setcounter{secnumdepth}{3}  
\setcounter{equation}{0}
\setcounter{figure}{0}
\renewcommand{\theequation}{S-\arabic{equation}}
\renewcommand{\thefigure}{S\arabic{figure}}
\renewcommand\figurename{Supplementary Figure}
\renewcommand\tablename{Supplementary Table}
\newcommand\Scite[1]{[S\citealp{#1}]}
\makeatletter \renewcommand\@biblabel[1]{[S#1]} \makeatother

\subsection*{Extraction of an effective  Hinshelwood cycle from a complex autocatalytic network}
Here we elucidate the application of Eq.~(11) of the main text through a specific example. Consider an $N=8$ SHC with two additional reactions (connections), (i) $X_{6} \eqwithrate{\a\,  x_{6}} X_{6} + X_{1}$ and (ii) $X_{1} \eqwithrate{\b\,  x_{1}} X_{1} + X_{5}$. Thus, the new reaction matrix, $\tilde{\mb{K}}$, has elements $\tilde{\mb{K}}_{1 6} = \a$, $\tilde{\mb{K}}_{51} = \b$ and all other elements of $\tilde{\mb{K}}$ are equal to the respective elements of the reaction matrix, $\mb{K}$,  of the $N=8$ SHC. For simplicity, we assume that all the original SHC rates are equal, and rescale all other rates (i.e., $\a$ and $\b$) by them, effectively setting the original SHC rates  to $1$. The equivalent network connectivity is shown in Supplementary Fig.~\ref{fig-K}(c). This network factorizes into $4$  cycles, as shown in Supplementary Fig.~\ref{fig-loops}. The geometric mean of the respective rates for each of these four constitutive cycles are, $\k_{8} = (1)^{1/8} = 1, \k_{6} = (\a)^{1/6}, \k_{5} = (\b)^{1/5}, \k_{3} = (\a\,\b)^{1/3}$, respectively for the $8, 6, 5$ and $3$ member cycles. (See Supplementary Fig.~\ref{fig-loops}.) Therefore, using  using Eq~9 of the main text, the characteristic polynomial that determines the eigenvalues of $\tilde{\mb{K}}$ is, 
%%%%%%%%%%%%%%%%%%%%%%%%begin%%%%%%%%%%%%%%%%%%%%%%%%
\begin{align}%\label{eq-}
1 &= \frac{1}{\l^8} + \frac{\a}{\l^6} +  \frac{\b}{\l^5} +  \frac{\a \b}{\l^3},  \nn\\
\implies \l^8 &= 1 + \a \, \l^2 + \b \, \l^3 + \a \b \, \l^5. 
 \end{align}
%%%%%%%%%%%%%%%%%%%%%%%%%end%%%%%%%%%%%%%%%%%%%%%%%%

Asymptotically, the eigenvalue that has  the largest positive real part dominates the overall dynamics. Using the above equation it can be shown that the asymptotic dynamics of the whole network entrains to the cycle with the largest $\k_{\tiny{cycle}}$, for a given value of $\a$ and $\b$.

%%%%%%%%%%%%%%%%%%%%%begin%%%%%%%%%%%%%%%%%%%%
\begin{figure}[h]
\begin{center}
\resizebox{16cm}{!}{\includegraphics{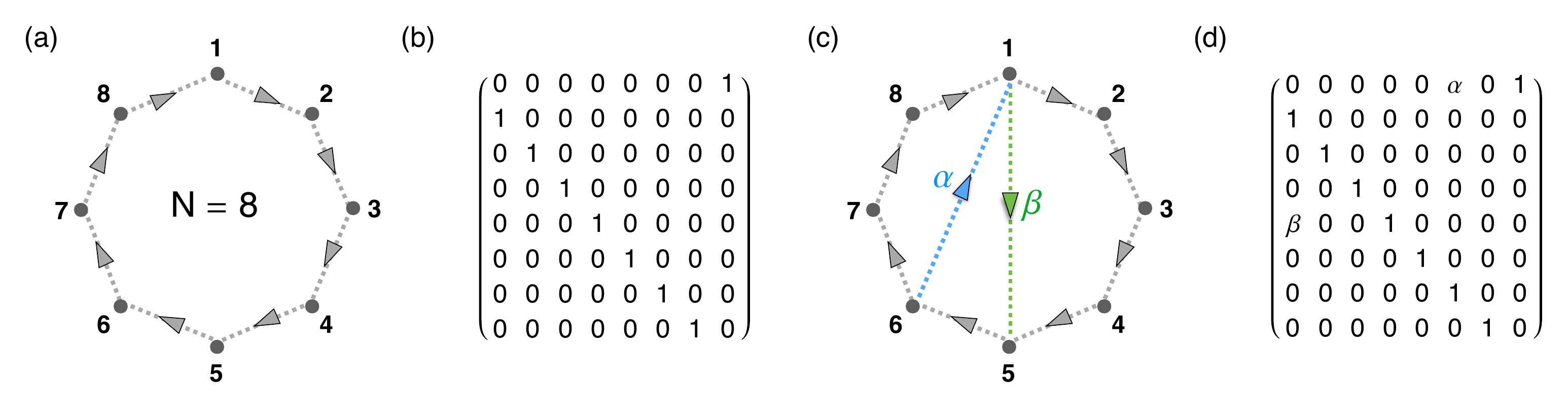}}%inclgphcs[trim=lcm bcm rcm tcm, clip=true, angle=-90]
\caption{The network connectivity of the $N=8$ stochastic Hinshelwood cycle (SHC) is shown in (a). Its reaction matrix, $\mb{K}$, is shown in (b); all SHC rates are assumed equal, and then set equal to $1$. In (c) we show an  $N=8$ SHC, with two extra connections, corresponding to the two reactions, $X_{6} \eqwithrate{\a\,  x_{6}} X_{6} + X_{1}$ (in blue) and $X_{1} \eqwithrate{\b\,  x_{1}} X_{1} + X_{5}$ (in green).  The reaction matrix for the network in (c), $\tilde{\mb{K}}$, is shown in (d).}
\label{fig-K}
\end{center}
\end{figure}
%%%%%%%%%%%%%%%%%%%%%%end%%%%%%%%%%%%%%%%%%%%

%%%%%%%%%%%%%%%%%%%%%begin%%%%%%%%%%%%%%%%%%%%
\begin{figure}[h]
\begin{center}
\resizebox{16cm}{!}{\includegraphics{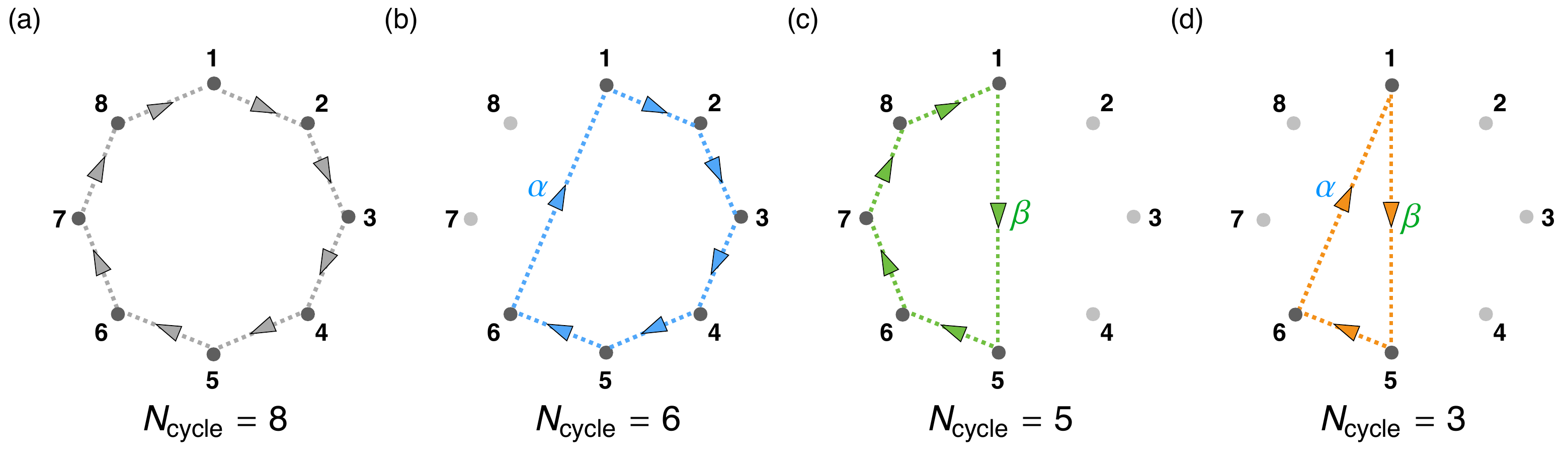}}%inclgphcs[trim=lcm bcm rcm tcm, clip=true, angle=-90]
\caption{The augmented $N=8$ SHC network shown in Supplementary Fig.~\ref{fig-K}~(c), with the reaction matrix, $\tilde{\mb{K}}$, shown in  Supplementary Fig.~\ref{fig-K}~(d), can be factored into four constituent cycles, shown here in (a), (b), (c) and (d). The total number of reactants implicated in each of these cycles, $N_{\tiny{cycle}}$, is also shown. The geometric mean of the rates for each of these cycles, $\k_{{\tiny{cycle}}}$, is therefore, (a) $\k_{8} = 1$, (b)  $\k_{6} = (\a)^{1/6}$, (c) $\k_{5} = (\b)^{1/5}$ and (d) $\k_{3} = (\a\,\b)^{1/3}$. 
}
\label{fig-loops}
\end{center}
\end{figure}
%%%%%%%%%%%%%%%%%%%%%%end%%%%%%%%%%%%%%%%%%%%

%%%%%%%%%%%%%%%%%Supplemental Material end%%%%%%%%%%%%%%%%%

\end{document}